\documentclass[a4,referee,pre,showpacs]{revtex4}

\usepackage{graphicx}
\usepackage{amsmath}
\usepackage{natbib}

\newcommand\EQ{\begin{equation}}
\newcommand\EN{\end{equation}}
\newcommand\EQA{\begin{eqnarray}}
\newcommand\ENA{\end{eqnarray}}

\begin{document}

\title{Thermodynamics of MHD flows with axial symmetry}
\author{N. Leprovost}
\email{nicolas.leprovost@cea.fr}
\affiliation{DRECAM/SPEC/CEA Saclay, and CNRS (URA2464), F-91190 Gif sur Yvette Cedex, France}
\author{P-H. Chavanis}
\affiliation{Laboratoire de Physique Th\'eorique (UMR 5152), Universit\'e Paul Sabatier, Toulouse, France}
\author{B. Dubrulle}
\affiliation{DRECAM/SPEC/CEA Saclay, and CNRS (URA2464), F-91190 Gif sur Yvette Cedex, France}

\begin{abstract}
We present strategies based upon extremization principles, in the case of the axisymmetric equations of magnetohydrodynamics (MHD).  
We study the equilibrium shape by using a minimum energy principle under the constraints of the MHD axisymmetric equations. 
We also propose a numerical algorithm based on a maximum energy dissipation principle to compute in a consistent way the equilibrium states. Then, we develop the statistical mechanics of such flows and recover the same equilibrium 
states giving a justification of the minimum energy principle. We find that fluctuations obey a Gaussian shape  
and we make the link between the conservation of the Casimirs on the coarse-grained scale and the process of energy dissipation.
\end{abstract}

\maketitle

\section{Introduction}
The recent success of two experimental fluid dynamos 
\cite{Gailitis01,Stieglitz01} has renewed the interest in the 
mechanism of dynamo saturation, and, thus, of equilibrium 
configurations in MHD. At the present time, there is no general theory 
to tackle this problem, besides dimensional theory. For example, in a 
conducting fluid with typical velocity $V$, density $\rho$ , Reynolds 
number $Re$ and magnetic Prandtl number $Pm$, the typical level of 
magnetic field reached at saturation is necessarily \cite{Petrelis01}:
\EQ
B^2=\mu_o\rho V^2 f(Re,Pm),
\label{saturation}
\EN
where $f$ is a priori an arbitrary function of $Re$ and $Pm$. Many 
numerical simulations \cite{Archontis00} lead to $f=1$, i.e. 
equipartition between the magnetic and turbulent energy. This is 
therefore often taken as a working tool in astrophysical or 
geophysical application. However, this result is far from applying to 
any saturated dynamo. Moreover, it does not give any information 
about possible anisotropy of the saturated field. It would therefore 
be interesting to build robust algorithms to derive the function $f$. 
By robust, we mean algorithm which depends on characteristic global 
quantities of the system (like total energy) but not necessarily on 
small-scale dissipation, or boundary conditions.\

An interesting candidate in this regards is provided by statistical 
mechanics. In the case of pure fluid mechanics, statistical mechanics 
has mainly been developed within the frame of Euler equation  for
a two-dimensional perfect fluid. Onsager \cite{Onsager49} first used 
a Hamiltonian model of point vortices. Within this framework, 
turbulence is a state of negative temperature leading to the 
coalescence of vortices of same sign \cite{Montgomery74}. Further 
improvement were provided by Miller et al. \cite{Miller90} and Robert 
and Sommeria \cite{Robert91} who independently introduced a 
discretization of the vorticity in a certain number of levels to
account for the continuous nature of vorticity. Using the maximum
entropy formalism of statistical mechanics \cite{Jaynes57}, it is then
possible to give the shape of the (meta)-equilibrium solution of
Euler's equation as well as the fine-grained fluctuations around it
\cite{jfm1}. This is similar to Lynden-Bell's theory of violent
relaxation
\cite{lb} in stellar dynamics (see Chavanis \cite{houches} for a
description of the analogy between 2D vortices and stellar
systems). The predictive power of the statistical theory is however
limited by the existence of an infinite number of constants (the
Casimirs) which precludes the finding of an universal Gibbs state. In particular, the metaequilibrium state strongly 
depends on the details of the initial condition.  In certain
occasions, for instance when the flow is forced at small scales, it may be
more relevant to fix a prior distribution of vorticity fluctuations
instead of the Casimirs \cite{ellis}. Then, the coarse-grained flow
maximizes a ``generalized'' entropy functional determined by the prior
distribution of vorticity \cite{Chavanis03}.  The statistical
mechanics of MHD flows has been recently explored by Jordan and
Turkington \cite{Jordan97} in 2D. In contrast with non-magnetized 2D
hydrodynamics they obtained a {\it universal} Gaussian shape for the
fluctuations. This comes from the fact that the conserved quantity in
the MHD case is an integral quantity of the primitive velocity and
magnetic fields and thus, in the continuum limit, has vanishing
fluctuations.\

The pure 2D situation however seldom applies to astrophysical or 
geophysical flows. In this respect, it is interesting to develop 
statistical mechanics of systems closer to natural situations, albeit 
sufficiently simple so that the already well tested recipes of 
statistical mechanics apply. These requirements are met by flows with 
axial symmetry. Most natural objects are rotating, selecting this 
peculiar symmetry. Moreover, upon shifting from 2D to axi-symmetric 
flows, one mainly shifts from a translation invariance along one 
axis, towards a rotation invariance along one axis. Apart from 
important physical consequences which need to be taken into account 
(for example conservation of angular momentum instead of vorticity or 
momentum, curvature terms), this induces a similarity between the two 
systems which enables a natural adaptation of the 2D case to the 
axisymmetric case. This is shown in the present paper, where we 
recover the Gaussian shape of the fluctuations and make the link 
between the conservation of the Casimirs on the coarse-grained scale 
and the process of energy dissipation.\

In the first part of the paper, we study the equilibrium shape by
using a minimum energy principle under the constraints of the MHD
axisymmetric equations. We also propose a numerical algorithm based on
a maximum energy dissipation principle to compute in a consistent way
the equilibrium states. This is similar to the relaxation equation
proposed by Chavanis \cite{Chavanis03} in 2D hydrodynamics to
construct stable stationary solutions of the Euler equation by
maximizing the production of a $H$-function.  Then, we develop the
statistical mechanics of such flows and recover these equilibrium
states, thereby providing a physical justification for the minimum
energy principle.

\section{MHD flows with axial symmetry}
\subsection{Equations and notations}
Consider the ideal incompressible MHD equations:
\EQA
\partial_t {\bf U}+({\bf U} \cdot \nabla){\bf U} &=& - \frac{1}{\rho} \nabla P
+({\bf \nabla}\times{\bf B})\times{\bf B} \, ,\nonumber\\
\partial_t {\bf B}+({\bf U} \cdot \nabla){\bf B}&=&({\bf B} \cdot 
\nabla){\bf U} \, ,
\label{idealMHD}
\ENA
where ${\bf U}$ is the fluid velocity, $P$ is the pressure, 
$\sqrt{\rho\mu_0} \, {\bf B}$ is the magnetic field and $\rho$ is the 
(constant) fluid density.
In the axisymmetric case we consider, it is convenient to introduce 
the poloidal/toroidal decomposition for the fields ${\bf U}$ and 
${\bf B}$:
\EQA
{\bf U}&=&{\bf U_p} + {\bf U_t} = {\bf U_p} + U \, {\bf e_{\theta}} \, ,
\\ \nonumber
{\bf B}&=&{\bf B_p} + {\bf B_t}= {\bf \nabla} \times (A {\bf 
e_{\theta}}) + B \, {\bf e_{\theta}} \, ,
\ENA
where ${\bf A}={\bf A_p} + A \, {\bf e_{\theta}}$ is the potential 
vector. This decomposition will be used in our statistical mechanics 
approach.\

When considering energy methods, we shall introduce alternate fields, 
built upon the poloidal and toroidal decomposition. They are : 
$\sigma_u = rU$,
$\sigma_b=rA$, $\xi_u=\omega/r$ and $\xi_b=B/r$, where $\omega$ is 
the toroidal part of the vorticity field. In these variables, the 
ideal incompressible MHD equations (\ref{idealMHD}) become, in the 
axisymmetric approximation, a set of four scalar equations:
\EQA
\label{MHDcyl}
\partial_t \sigma_b + \{\psi,\sigma_b\} &=& 0 \, , \\ \nonumber
\partial_t \xi_b + \{\psi,\xi_b\} &=& 
\{\sigma_b,\frac{\sigma_u}{2y}\} \, , \\ \nonumber
\partial_t \sigma_u + \{\psi,\sigma_u\} &=& \{\sigma_b, 2y \xi_b\}  \, , \\ \nonumber
\partial_t \xi_u + \{\psi,\xi_u\} &=& \partial_z 
(\frac{\sigma_u^2}{4y^2}-\xi_b^2)-\{\sigma_b, \Delta_{*} \sigma_b\}  \, , 
\ENA
where the fields are function of the axial coordinate $z$ and the 
modified radial coordinate $y=r^2/2$ and $\psi$ is a stream function: 
${\bf U_p} = {\bf \nabla} \times (\psi / r \quad {\bf e_{\theta}})$. 
We have introduced a Poisson Bracket: $\{f,g\}= \partial_y f 
\partial_z g - \partial_z f \partial_y g$. We also defined a pseudo 
Laplacian in the new coordinates:
\EQ
\Delta_{*}=\frac{\partial^{2}}{\partial 
y^{2}}+\frac{1}{2y}\frac{\partial^{2}}{\partial z^{2}} \, .
\EN
Following \cite{Jordan97}, we will make an intensive use of the 
operators (for more details, see appendix \ref{AnnexeA}): $curl$ 
which gives the toroidal part of the curl of any vector and ${\bf 
Curl}$ which takes a toroidal field as argument and returns the 
poloidal part of the curl.
If $j = curl {\bf B}$ is the toroidal part of the current and $\psi = 
r \, Curl^{-1}({\bf U_p})$, the following relations hold:
\EQA
\label{xi-u}
\xi_u &=& - \Delta_{*} \psi  \, ,  \\
\label{courant}
j/r &=& - \Delta_{*} \sigma_b \, .
\ENA
Under the shape (\ref{MHDcyl}), the ideal axisymmetric MHD equation 
of motion lead to the immediate identification of $\sigma_b = rA$ as 
a conserved quantity associated to axial symmetry. This quantity is 
only advected by the velocity field and thus should play a special 
role regarding the global conserved quantities, as we now show.

\subsection{Conservation laws}
\subsubsection{General case}
The whole set of conservation laws of the axisymmetric ideal MHD 
equations have been derived by Woltjer \cite{Woltjer59}:
\EQA
\label{cons1}
I&=&\int C(\sigma_{b}) \, dydz  \, , \\ \nonumber
H_{m}&=&2\int\xi_{b}N(\sigma_{b}) \, dydz  \, , \\ \nonumber
H_{c}&=&\int \lbrace 
F(\sigma_{b})\xi_{u}+\sigma_{u}\xi_{b}F'(\sigma_{b})\rbrace  \, dydz  \, , 
\\ \nonumber
L&=&\int\sigma_{u}G(\sigma_{b}) \, dydz  \, , \\ \nonumber
E&=&\frac{1}{2}\int \biggl\lbrace 
\xi_{u}\psi-\sigma_{b}\Delta_{*}\sigma_{b} + 
\frac{\sigma_{u}^{2}}{2y} + 2y\xi_{b}^{2}\biggr\rbrace \,  dydz  \, .
\ENA
where $C$, $N$, $F$ and $G$ are arbitrary functions. One can check 
that these integrals are indeed constants of motion by using 
(\ref{MHDcyl}) and the following boundary conditions: $\sigma_b = 
\sigma_u = \xi_u =\xi_b =0$ on the frontier of the domain. To prove 
the constancy of the third integral, one has to suppose that 
$F(0)=0$. The interpretation of these integrals of motion is easier 
when considering a special case, introduced by Chandrasekhar 
\cite{Chandrasekhar58}.

\subsubsection{Chandrasekhar model}

The conservation laws take a simpler shape when one considers only 
linear and quadratic conservation laws, such that 
\cite{Chandrasekhar58} 
$N(\sigma_{b})=F(\sigma_{b})=G(\sigma_{b})=\sigma_{b}$ and 
$N(\sigma_b) = G(\sigma_b) = 1$ . The case $F(\sigma_b) = 1$ is 
forbidden by the condition that $F$ should vanish at the origin. In 
that case, the set of conserved quantities can be split in two families. The first one is made-up with conserved quantities of the ideal MHD system, irrespectively of the geometry:
\EQA
\label{cons2}
H_{m} &=& 2\int\xi_{b}\sigma_{b}  \, dydz =\int {\bf A}\cdot {\bf B} 
\,  d{\bf x}=2\int A B \,  d{\bf x} \, , \\ \nonumber
H_{c} &=&\int \lbrace \sigma_{b}\xi_{u}+\sigma_{u}\xi_{b}\rbrace \, 
dydz= \int {\bf U}\cdot {\bf B}  \, d{\bf x} \, , \\ \nonumber
E &=&\frac{1}{2}\int \biggl\lbrace 
\xi_{u}\psi-\sigma_{b}\Delta_{*}\sigma_{b} + 
\frac{\sigma_{u}^{2}}{2y} + 2y\xi_{b}^{2}\biggr\rbrace \,  dydz=
\frac{1}{2}\int ({\bf U}^{2}+{\bf B}^{2}) \,  d{\bf x} \, , 
\ENA
where $H_m$ is the magnetic helicity, $H_c$ is the cross-helicity and $E$ is the total
energy. Note that due to the Lorentz force, the kinetic helicity is
not conserved, unlike in the pure hydrodynamical case. The other family of conserved quantities is made of the particular integrals of motion which appear due to {\em axisymmetry}:
\EQA
\label{cons3}
I &=&\int C(\sigma_{b})dydz= \int C(rA) \, d{\bf x} \, , \\ \nonumber
H_{m}' &=& 2\int\xi_{b} \, dydz =\int \frac{B}{r} d{\bf x} \, , \\ \nonumber
L&=&\int\sigma_{u}G(\sigma_{b}) \, dydz  = \int r^2 U B \, d{\bf x} \, , \\ \nonumber
L'&=&\int\sigma_{u} \, dydz = \int r U  \, d{\bf x} \, .
\ENA
Apart from $L'$ the angular momentum, it is difficult to give the other quantities any physical interpretation. The class of invariant $I$ are called the Casimirs of the system (if one defines a non canonical bracket for the Hamiltonian system, they commute, in the bracket sense, will all other functionals). The conservation laws found by Woltjer are then generalization of these quantities.

\subsection{Dynamical stability}
\label{EnergyMethod}
\subsubsection{General case}

Following \cite{Woltjer59}, we show that the extremization of energy at
fixed $I$, $H_{m}$, $H_{c}$ and $L$ determines the general form of
stationary solutions of the MHD equations. We argue that the solutions
that {\it minimize} the energy are nonlinearly dynamically stable for
the inviscid equations.

To make the minimization, we first note that each integral is equivalent to an infinite set of constraints. Following Woltjer, we shall introduce a complete set of functions and label these functions and the corresponding integrals with an index $n$. Then, introducing Lagrange multipliers for each constraint, to first order, the variational problem takes the form:
\EQ
\delta E + \sum_{n=1}^{+\infty}\biggl\lbrace \alpha^{(n)}\delta I^{(n)} + \mu_{m}^{(n)} \delta H_m^{(n)} + \mu_{c}^{(n)}\delta H_c^{(n)} + \gamma^{(n)}
\delta L^{(n)}\biggr\rbrace = 0 \, .
\label{variation1}
\EN
Taking the variations on $\sigma_{b}$, $\xi_{b}$, $\sigma_{u}$ and $\xi_{u}$, we find:
\EQA
\label{Equilibr1}
\Delta_{*}\sigma_{b} &=& - F'(\sigma_{b})\Delta_{*}\psi + 
F''(\sigma_{b})\sigma_{u}\xi_{b} + G'(\sigma_{b})\sigma_{u} + 
2N'(\sigma_{b})\xi_{b}+C'(\sigma_{b}) \, ,
\\ \nonumber
2y\xi_{b} &=& -2N(\sigma_{b}) - F'(\sigma_{b})\sigma_{u} \, ,
\\ \nonumber
\frac{\sigma_{u}}{2y} &=& - F'(\sigma_{b})\xi_{b} - G(\sigma_{b}) \, ,
\\ \nonumber
\psi &=& - F(\sigma_{b}) \, ,
\ENA
where we have set
$F(\sigma_{b})=\sum_{n=0}^{+\infty}\mu_{c}^{(n)}F_{n}(\sigma_{b})$ and
similar notations for the other functions.  This is the general
solution of the incompressible axisymmetric ideal MHD problem
\cite{Woltjer59}. In the general case, it is possible to express the
three field $\sigma_u$, $\xi_u$ and $\xi_b$ in terms of
$\sigma_b$. Then the first equation of the above system leads a
partial differential equation for $\sigma_b$ to be solved to find the
equilibrium distribution. Note that the extremization of the
``free energy'' $J=E + \alpha I + \mu_m H_m + \mu_c H_c + \gamma L$
yields the same equations as (\ref{Equilibr1}). Differences will appear
on the second order variations (see below).

\subsubsection{Chandrasekhar model}

In the Chandrasekhar model, the arbitrary functions are at most linear
functions of $\sigma_b$: $N(\sigma_{b})=\mu_m
\sigma_{b}+\mu'_m$, $F(\sigma_{b})=\mu_c \sigma_{b}$ and 
$G(\sigma_{b})=\gamma\sigma_{b}+\gamma'$. Thus the stationary
profile in the Chandrasekhar model is given by:
\EQA
\label{Equilibr2}
\Delta_{*}\sigma_{b} &=& - \mu_c \Delta_{*}\psi + \gamma \sigma_{u} + 
2 \mu_m \xi_{b} + C'(\sigma_{b}) \, ,
\\ \nonumber
2y \xi_{b} &=& -2 \mu_m \sigma_{b} - 2 \mu'_m - \mu_c \sigma_{u} \, ,   
\\ \nonumber
\frac{\sigma_{u}}{2y} &=& - \mu_c \xi_{b} - \gamma \sigma_{b} - \gamma' \, ,
\\ \nonumber
\psi &=& -\mu_c \sigma_{b} \, .
\ENA
 From the previous equations, we obtain
\EQA
\label{Eqn}
2 y (1-\mu_c^{2}) \xi_{b} &=& 2(\gamma \mu_c y - \mu_m) \sigma_{b} + 
2 \mu_c \gamma' y -2\mu'_m \, ,
\\ \nonumber
(1 - \mu_c^{2}) \sigma_{u} &=& 2(\mu_c \mu_m - \gamma y) \sigma_{b} + 
2 \mu_c \mu'_m - 2\gamma' y \, ,
\\ \nonumber
\psi &=& -\mu_c \sigma_{b} \, ,
\ENA
where $\sigma_{b}$ is given by the differential equation:
\begin{equation}
(1-\mu_c^{2})^{2} \Delta_{*}\sigma_{b} = \Phi(\sigma_{b}) - [ 
2\mu_m^{2} \frac{\sigma_{b}}{y} + 2\gamma^{2} y] \sigma_{b} - 2 
\gamma \gamma' y - \frac{2\mu_m \mu'_m}{y} \, .
\label{Equasigb}
\end{equation}
These expressions can be used to prove that these fields are
stationary solutions of the axisymmetric MHD equations. We now turn to
the stability problem. Since the free energy $J=E + \alpha I +
\mu_m H_m + \mu_c H_c + \gamma L$ is conserved by the ideal dynamics,
a minimum of $J$ will be nonlinearly dynamically stable (at least in
the {\it formal} sense of Holm et al. \cite{Holm85}).  Note that this
implication is not trivial because the system under study is
dimensionally infinite. We will admit that their analysis can be
generalized to the axisymmetric case. Since the integrals which appear
in the free energy are conserved individually, a minimum of energy at
fixed other constraints also determines a nonlinearly dynamically
stable stationary solution of the MHD equations. This second stability
criterion is stronger than the first (it includes it). We shall not prove these results, nor write the second order variations, here.

\subsection{Numerical algorithm}
\subsubsection{General case}

It is usually difficult to solve directly the system of equations (\ref{Eqn})-(\ref{Equasigb}) and make sure that they
yield a stable stationary solution of the MHD equations. Instead, we shall propose a set of relaxation equations which minimize the energy while conserving any other integral of motion. This permits to construct solutions of the system (\ref{Eqn})-(\ref{Equasigb}) which are energy minima and respect the other constraints. A physical justification of this precedure linked to the dissipation of energy will be given in Sec. \ref{Classical}. 

Our relaxation equations can be written under the generic form
\EQ
\label{relaxeq}
\frac{\partial \sigma}{\partial t} = -\nabla \cdot {\bf J}_{\sigma},
\EN
where $\sigma$ stands for $\sigma_{b}$, $\xi_{b}$, $\sigma_{u}$ or 
$\xi_{u}$. Using straightforward integration by parts, we then get:
\EQA
\label{contr2}
\dot I &=& \int {\bf J}_{\sigma_{b}}\cdot \lbrack \nabla 
C'(\sigma_{b}) \rbrack dydz \, ,
\\ \nonumber
\dot H_{m} &=& 2\int\biggl\lbrace {\bf J}_{\xi_{b}}\cdot \nabla 
\lbrack N(\sigma_{b})\rbrack +{\bf J}_{\sigma_{b}}\cdot \nabla\lbrack 
N'(\sigma_{b})\xi_{b}\rbrack\biggr\rbrace dydz \, ,
\\ \nonumber
\dot H_{c} &=& \int\biggl\lbrace {\bf J}_{\xi_{u}}\cdot \nabla 
\lbrack F(\sigma_{b})\rbrack +{\bf J}_{\sigma_{b}}\cdot \nabla\lbrack 
F'(\sigma_{b})\xi_{u}+F''(\sigma_{b})\sigma_{u}\xi_{b}\rbrack \\ 
\nonumber
&+& {\bf J}_{\sigma_{u}}\cdot \nabla\lbrack 
F'(\sigma_{b})\xi_{b}\rbrack +{\bf J}_{\xi_{b}}\cdot \nabla\lbrack 
F'(\sigma_{b})\sigma_{u}\rbrack \biggr\rbrace dydz \, ,
\\ \nonumber
\dot L &=& \int\biggl\lbrace {\bf J}_{\sigma_{u}}\cdot \nabla \lbrack 
G(\sigma_{b})\rbrack +{\bf J}_{\sigma_{b}}\cdot \nabla\lbrack 
G'(\sigma_{b})\sigma_{u}\rbrack\biggr\rbrace dydz \, ,
\\ \nonumber
\dot E &=& \int\biggl\lbrace {\bf J}_{\xi_{u}}\cdot \nabla \psi -{\bf 
J}_{\sigma_{b}}\cdot \nabla (\Delta_{*}\sigma_{b})
+{\bf J}_{\sigma_{u}}\cdot \nabla \biggl (\frac{\sigma_{u}}{2y}\biggr 
)+{\bf J}_{\xi_{b}}\cdot \nabla (2y\xi_{b}) \biggr\rbrace dydz \, .
\ENA
To construct the optimal currents, we rely on a procedure of maximization of the rate of dissipation of energy $\dot E$  very similar to the procedure of maximum entropy production principle (MEPP) of Robert and Sommeria \cite{Robert92} in the 2D turbulence case. This is equivalent to say that the evolution towards the equilibrium state (\ref{Eqn})-(\ref{Equasigb}) is very rapid. We thus try to maximize $\dot E$  given the conservation of $\dot I$, $\dot H_{m}$, $\dot H_{c}$ and $\dot L$. Such maximization can only have solution for bounded currents (if not, the fastest evolution is for infinite currents). Therefore, we also impose a bound on $J_{\sigma}^{2}$ where, as before, $\sigma$ stands for $\sigma_{b}$, $\xi_{b}$, $\sigma_{u}$, $\xi_{u}$.

Writing the variational problem under the form
\begin{eqnarray}
\delta \dot E + \sum_{n=1}^{+\infty}\biggl\lbrace \alpha^{(n)}(t)\delta \dot{I}^{(n)} + \mu_{m}^{(n)}(t) \delta \dot H_m^{(n)} + \mu_{c}^{(n)}(t)\delta \dot H_c^{(n)} + \gamma^{(n)}(t)\delta \dot L^{(n)}\biggr\rbrace -\sum_{\sigma}\frac{1}{D_{\sigma}}\delta\biggl 
(\frac{J_{\sigma}^{2}}{2}\biggr )=0
\end{eqnarray}
and taking variations on ${\bf J}_{\sigma_{b}}$, ${\bf J}_{\xi_{b}}$,
${\bf J}_{\sigma_{u}}$, ${\bf J}_{\xi_{u}}$, we obtain the optimal 
currents. Inserting their expressions in the relaxation equations 
(\ref{relaxeq}), we get:
\EQA
\frac{\partial\sigma_{b}}{\partial t}&=&\nabla\cdot \biggl\lbrace 
D_{\sigma_{b}}\nabla\cdot \biggl\lbrack 
-\Delta_{*}\sigma_{b}+C'(\sigma_{b},t)+2\xi_{b}N'(\sigma_{b},t) \\ 
\nonumber
&+&\xi_{u}F'(\sigma_{b},t)+\sigma_{u}\xi_{b}F''(\sigma_{b},t)+G'(\sigma_{b},t)\sigma_{u}\biggr\rbrack \biggr\rbrace \, ,
\\ \nonumber
\frac{\partial\xi_{b}}{\partial t}&=&\nabla\cdot \biggl\lbrace 
D_{\xi_{b}}\nabla\cdot \biggl\lbrack 
2y\xi_{b}+2N(\sigma_{b},t)+F'(\sigma_{b},t)\sigma_{u}\biggr\rbrack 
\biggr\rbrace \, ,
\\ \nonumber
\frac{\partial\sigma_{u}}{\partial t}&=&\nabla\cdot \biggl\lbrace 
D_{\sigma_{u}}\nabla\cdot \biggl\lbrack 
\frac{\sigma_{u}}{2y}+\xi_{b}F'(\sigma_{b},t)+G(\sigma_{b},t)\biggr\rbrack 
\biggr\rbrace \, ,
\\ \nonumber
\frac{\partial\xi_{u}}{\partial t}&=&\nabla\cdot \biggl\lbrace 
D_{\xi_{u}}\nabla\cdot \lbrack \psi+F(\sigma_{b},t)\rbrack 
\biggr\rbrace.
\ENA
where we have set $F(\sigma_{b},t)=\sum_{n=0}^{+\infty}\mu_{c}^{(n)}(t)F_{n}(\sigma_{b})$
and similar notations for the other functions. The time evolution of
the Lagrange multipliers $\mu_{c}^{(n)}(t)$ etc. are obtained by
substituting the optimal currents in the constraints $\dot
H_{c}^{(n)}=0$ etc. and solving the resulting set of algebraic equations.
Using the expression of the optimal currents and the condition that
$\dot I = \dot H_m = \dot H_c = \dot L = 0$, we can show that:
\begin{eqnarray}
\dot E=-\int\biggl\lbrace \frac{J_{\xi_{u}}^{2}}{ D_{\xi_{u}}}+ 
\frac{J_{\sigma_{b}}^{2}}{ D_{\sigma_{b}}}+ 
\frac{J_{\sigma_{u}}^{2}}{ D_{\sigma_{u}}}+ \frac{J_{\xi_{b}}^{2}}{ 
D_{\xi_{b}}}\biggr \rbrace dydz\le 0,
\end{eqnarray}
provided that the diffusion currents $D_{\xi_{u}}$, $D_{\sigma_{b}}$,$D_{\sigma_{u}}$ and $D_{\xi_{b}}$ are positive. 
Thus, the energy decreases until all the currents vanish. In that case, we obtain the static equations (\ref{Equilibr1}). In addition, this numerical algorithm guarantees that only energy minima (not maxima or saddle points) are reached. Note that if we fix the Lagrange multipliers instead of the constraints, the foregoing relaxation equations lead to a stationary state which minimizes the free energy $J$. Then, as stated above, the constructed solutions will be nonlinearly dynamical stable solution of the MHD set of equations. However, not allowing the Lagrange multiplier to depend on time, we may "miss" some stable solutions of the problem. Indeed, we know that minima of the free energy are nonlinearly stable solutions of the problem but we do not know if they are the only ones: some solutions can be minima of $E$ at fixed $I$, $H_{m}$, $H_{c}$ and $L$ while they are not minima of $J=E+\alpha I+\mu_{m}H_{m}+\mu_{c}H_{c}+\gamma L$. 

\subsubsection{Chandrasekhar model}
In the Chandrasekhar model (with $\mu'_m = \gamma' = 0$), the previous equations can be simplified.  The equilibrium solution does not depend on the particular value of the diffusion coefficients (these are only  multiplicative factors of the optimal currents) and for simplicity,  we set $D_{\xi_{u}}=D_{\sigma_{b}}=D_{\sigma_{u}}=D_{\xi_{b}}=1$. The relaxation equations then reduce to:
\EQA
\frac{\partial\sigma_{b}}{\partial t} &=& \Delta \biggl\lbrace 
-\Delta_{*}\sigma_{b}+ C'(\sigma_{b},t) + 2\mu_m(t)\xi_{b} + 
\mu_c(t)\xi_{u} + \gamma(t)\sigma_{u}\biggr\rbrace \, ,
\\ \nonumber
\frac{\partial\xi_{b}}{\partial t} &=& \Delta \biggl\lbrace 2y\xi_{b} 
+ 2\mu_m(t)\sigma_{b}+\mu_c(t)\sigma_{u}\biggr\rbrace \, ,
\\ \nonumber
\frac{\partial\sigma_{u}}{\partial t} &=& \Delta \biggl\lbrace 
\frac{\sigma_{u}}{2y} + \mu_c(t)\xi_{b}+\gamma(t)\sigma_{b} 
\biggr\rbrace \, ,
\\ \nonumber
\frac{\partial\xi_{u}}{\partial t} &=& \Delta \lbrace \psi + 
\mu_c(t)\sigma_{b}\rbrace.
\ENA
where the Lagrange multipliers evolve in time so as to conserve the 
constraints (\ref{contr2}).

These equations are the MHD counterpart of the relaxation equations
proposed by Chavanis \cite{Chavanis03} for 2D hydrodynamical flows
described by the Euler equation. In this context, a stable stationary
solution of the Euler equation maximizes a H-function (playing the
role of a generalized entropy) at fixed energy and circulation. A
justification of this procedure, linked to the increase of H-functions
on the coarse-grained scale, will be further discussed in Sec. \ref{Conclusion} and
compared with the MHD case.

If we set the velocity field to zero ($\sigma_u = \xi_u =0$), we get a
system of equations linking the poloidal part ($\sigma_b$) and the
toroidal part ($\xi_b$) of the magnetic field. It is fairly easy to
see that the coupling between the two quantities is proportional to
$\mu_m$ the Lagrange multiplier associated to the conservation of
magnetic helicity. This is reminiscent of the $\alpha$ effect of
dynamo theory (see Steenbeck et al. \cite{Steenbeck66}): in the
``kinematic approximation" where the effect of the Lorentz force is
removed, the coupling between the toroidal and poloidal part of the
magnetic field is given by a coefficient proportional to the kinetic
helicity of the fluctuating velocity field. Our model is not able to
recover this fact because, as noticed above, this quantity is not
conserved in the full MHD case. However, taking into account the
retroaction of the magnetic field on the velocity field, Pouquet et
al. \cite{Pouquet76} were able to write the non-linear $\alpha$-effect
as a difference between the kinetic and the magnetic helicity of the
fluctuations: $\alpha = H_k - H_m$. Our relaxation equations therefore recover
the fact that the approach to saturation of the magnetic field is
mainly monitored by the magnetic helicity.

\section{Statistical mechanics of axisymmetric flows}

In the previous section, we obtained general equilibrium velocity and magnetic field {\sl profiles} through minimization of the energy under constraints. In the present section, we derive velocity and magnetic field {\sl distribution} using
a thermodynamical approach, based upon a statistical mechanics of axisymmetric MHD flows. As we later check, the distribution we find are such that their mean fields obey the equilibrium profiles found by 
energy minimization. For simplicity, we focus here on the Chandrasekhar model. 

\subsection{Definitions and formalism}

Following \cite{Miller90}, \cite{Robert91} and \cite{Jordan97}, we
introduce a coarse-graining procedure through the consideration of a
length-scale under which the details of the fields are irrelevant.
The microstates are defined in terms of all the microscopic possible
fields ${\bf u}({\bf x})$ and ${\bf b}({\bf x})$. On this phase space,
we define the probability density $\rho({\bf r},{\bf u},{\bf b})$ of a
given microstate. The macrostates are then defined in terms of fields
observed on the coarse-grained scale. The mean field (denoted by a
bar) is determined by the following relations:
\EQA
\bar{\bf U}({\bf x}) = \int {\bf u} \, \rho({\bf r},{\bf u},{\bf b}) 
\, d{\bf u} d{\bf b} \, , \\ \nonumber
\bar{\bf B}({\bf x}) = \int {\bf b} \, \rho({\bf r},{\bf u},{\bf b}) 
\, d{\bf u} d{\bf b}\, .
\ENA
We introduce the mixing entropy
\EQ
S[\rho] = - \int \rho({\bf r},{\bf u},{\bf b}) \ln[\rho({\bf r},{\bf 
u},{\bf b})] \, d{\bf r} d{\bf u} d{\bf b} \, ,
\EN
which has the form of Shanon's entropy in information theory
\cite{Shanon49} \cite{Jaynes57}. The most probable states are the 
field $\overline{\bf U}$ and $\overline{\bf B}$ which maximize the
entropy subject to the constraints. The mathematical ground for such a
procedure is that an overwhelming majority of all
the possible microstates with the correct values for the constants of
motion will be close to this state (see \cite{Robert91} for
a precise definition of the neighborhood of a macrostate and the proof
of this concentration property). Note that this approach gives not only the coarse-grained field  ($\overline{\bf U}$, $\overline{\bf B}$) but also the fluctuations around it through the distribution $\rho({\bf r},{\bf u},{\bf b})$.

Each conserved quantity has a numerical value which can be
calculated given the initial condition, or from the detailed knowledge
of the fine-grained fields. The integrals calculated with the
coarse-grained quantities are not necessarily conserved because part
of the integral of motion can go into fine-grained fluctuations (as we
shall see, this is the case for the energy in MHD flows).  This
induces a distinction between two classes of conserved quantities,
according to their behavior through coarse-graining. Those which are
not affected are called robust, whereas the other one are called
fragiles.

\subsection{Constraints}

In this section, it is convenient to come back to the original 
velocity and magnetic fields. The constraints are the coarse-grained values of the conserved quantities 
(\ref{cons2}).
The key-point, as noted by \cite{Jordan97}, is that the 
quantity coming from a spatial integration of one of the field $\bf 
u$ or $\bf b$, is smooth. In our case, it amounts to neglecting the 
fluctuations of $A$ which is spatially integrated from ${\bf B}$ and 
write $A = \bar{A}$. Thus, the coarse-grained values of the conserved 
quantity are given by:
\EQA
\bar{I} &=& \int C(r\bar{A}) \, d{\bf x} \, , 
\\ \nonumber
\bar{H}_{m} &=& 2\int \bar{A}\ \bar{B} \, d{\bf x}  \, , 
\\ \nonumber
\bar{H}_{c} &=& \int {\bf u}\cdot {\bf b}\ \rho({\bf r},{\bf u},{\bf 
b}) \, d{\bf x} d{\bf u}d{\bf b}  \, , 
\\ \nonumber
\bar{E} &=&\frac{1}{2}\int ({\bf u}^{2}+{\bf b}^{2})\ \rho({\bf 
r},{\bf u},{\bf b})\ d{\bf x} \, d{\bf u}d{\bf b}  \, , \nonumber\\
\bar{H}_{m}'&=&2\int \frac{\bar{B}}{r}\ d{\bf x} \, ,
\\ \nonumber
\bar{L}&=&\int \bar{A}\bar{U}r^{2}\ d{\bf x} \, ,
\\ \nonumber
\bar{L}'&=&\int \bar{U}r\ d{\bf x} \, .
\ENA
The constraint $\bar{I}$ is the Casimir, connected to the conservation
of $\sigma_b$ along the motions. In the present case, it is a robust
quantity as it is conserved on the coarse-grained scale.  As stated
previously, the quantities $\bar{H}_{m}$, $\bar{H}_{c}$ and $\bar{E}$
are the mean values of the usual quadratic invariants of ideal MHD,
namely the magnetic helicity, the cross-helicity and the energy. On
the contrary, the quantities $\bar{H}_{m}'$, $\bar{L}$ and $\bar{L}'$
are specific to axisymmetric systems. Because these last three
conservation laws are usually disregarded in classical MHD
theory, it is interesting in the sequel to separate the study in two
cases, according to which the conservation of $\bar{H}_{m}'$,
$\bar{L}$ and $\bar{L}'$ is physically relevant (``rotating case") or
is not physically relevant (``classical case").\

\subsection{Gibbs state}
\subsubsection{Classical case}
\label{Classical}
The MHD equations develop a mixing process leading to a 
metaequilibrium state on the coarse-grained scale. It is obtained by 
maximizing the mixing entropy $S[\rho]$ with respect to the 
distribution $\rho$ at fixed $\bar{I}$, $\bar{H}_{m}$, $\bar{H}_{c}$ 
and $\bar{E}$ (we omit the bars in the following). We have:
\EQA
\delta S &=& -\int (1+\ln\rho)\ \delta\rho \, d{\bf x} d{\bf u}d{\bf b}  \, , 
\\ \nonumber
\delta {H}_{c} &=& \int {\bf u}\cdot {\bf b}\ \delta\rho \, d{\bf 
x} d{\bf u}d{\bf b} \, , 
\\ \nonumber
\delta {E} &=& \frac{1}{2}\int ({\bf u}^{2}+{\bf b}^{2})\ 
\delta\rho \, d{\bf x} d{\bf u}d{\bf b} \, .
\ENA
The variation of the magnetic helicity and the Casimirs is more 
tedious because they involve the coarse-grained field $\bar{A}$. For 
the magnetic helicity, we have:
\EQ
\delta H_{m}=2\int (\delta\bar{A}\ \bar{B}+\bar{A}\ \delta\bar{B})d{\bf x} \, .
\EN
Now, using an integration by parts, it is straightforward to show that
\EQ
\int \delta {A}\ {B}\ d{\bf x}= \int \delta{\bf B}_{P}\cdot {\bf 
A}_{P}\ d{\bf x} \, .
\EN
Therefore,
\EQA
\delta H_{m} &=& 2\int (\delta \bar{{\bf B}}_{P}\cdot \bar{{\bf 
A}}_{P}+\bar{A}\ \delta\bar{B})\ d{\bf x}=2\int \bar{\bf A}\cdot 
\delta\bar{\bf B}\ d{\bf x} \\ \nonumber
&=& 2\int \bar{\bf A}\cdot {\bf b}\ \delta\rho \ d{\bf x} d{\bf u}d{\bf b} \, .
\ENA
Regarding the variation of the Casimirs, we find:
\EQA
\delta I = \int C'(r\bar A)r\delta\bar A \, d{\bf x} = \int C'(r\bar 
A)r \ {\rm Curl}^{-1} \bar{\bf B}_{P} \ d{\bf x} = \int {\bf 
curl}^{-1}\lbrack rC'(r\bar A)\rbrack \cdot \delta \bar{\bf B}_{P} \ 
d{\bf x} \, ,
\ENA
or
\EQ
\delta I=\int {\bf curl}^{-1}\lbrack rC'(r\bar A)\rbrack \cdot {\bf 
b}_{P}\ \delta\rho \ d{\bf x} d{\bf u}d{\bf b} \, .
\EN

Writing the variational principle in the form
\EQ
\delta S-\beta\delta E-\mu_{m}\delta H_{m}-\mu_{c}\delta H_{c}-\sum_{n=1}^{+\infty}\alpha^{(n)}\delta I^{(n)} = 0 \, ,
\EN
we find that
\EQ
\label{g1}
1+\ln\rho= -\frac{\beta}{ 2}({\bf u}^{2}+{\bf 
b}^{2})-{2\mu_{m}}\bar{\bf A}\cdot {\bf b}-{\mu_{c}}{\bf u}\cdot {\bf 
b}-{\bf curl}^{-1}\lbrack rC'(rA)\rbrack\cdot {\bf b}_{P} \, .
\EN
It is appropriate to write ${\bf u}=\bar{\bf U}+{\bf u}'$ and ${\bf 
b}=\bar{\bf B}+{\bf b}'$ where the first term denotes the 
coarse-grained field. Then, the equation (\ref{g1}) can be
rewritten
\EQA
1+\ln\rho &=& - \frac{\beta}{2}(u'^2+b'^2) - \mu_c {\bf u'} \cdot 
{\bf b'} - \mu_m \bar{\bf A} \cdot \bar{\bf B} - \frac{\mu_c}{2} 
\bar{\bf U} \cdot \bar{\bf B}\\ \nonumber
&-& (\frac{\bar{\bf U}}{2}+ {\bf u'}) \cdot [\beta \bar{\bf U} + 
\mu_c \bar{\bf B}] \\ \nonumber
&-& ({\frac{\bar{\bf B}}{2} + \bf b'}) \cdot [\beta \bar{\bf B} + 
2\mu_m \bar{\bf A} + \mu_c \bar{\bf U} + {\bf curl}^{-1}\lbrack 
rC'(rA)\rbrack \, .
\ENA
Hence the fluctuations are Gaussian:
\EQ
\rho=\frac{1}{Z}\ {\rm exp}\biggl\lbrace - \frac{\beta}{2}({\bf 
u}'^{2}+{\bf b}'^{2})-{\mu_{c}}{\bf u}'\cdot {\bf b}'\biggr\rbrace
\label{g2}  = \frac{1}{Z}{\rm exp}\biggl\lbrace \frac{1}{2} \sum_{i,j} x_i A_{ij} x_j \biggr\rbrace \; ,
\EN
where we defined a 6-dimensionnal vector: $x_i = (u'_1,u'_2,u'_3,b'_1,b'_2,b'_3)$. The mean-field is given by:
\EQA
\beta {\bf U}+\mu_{c}{\bf B}={\bf 0} \, ,\\ \nonumber
\beta B+2\mu_{m}A+\mu_{c}U=0 \, ,\\ \nonumber
\beta{\bf B}_{P}+2\mu_{m}{\bf A}_{P}+\mu_{c}{\bf U}_{P}+{\bf 
curl}^{-1}\lbrack rC'(rA)\rbrack=0 \, .
\ENA
Taking the ${\rm curl}$ of these relations and using ${\rm curl}{\bf 
B}_{P}=j$, ${\rm curl}{\bf U}_{P}=\omega$ and ${\rm curl}{\bf 
A}_{P}=B$, we recover the equilibrium distribution (\ref{Equilibr2}) 
with $\gamma = \gamma' = \mu_m' = \mu_c' = 0$. Therefore, in this classical case, the equilibrium profiles are such 
that mean velocity and mean magnetic field are aligned. This is a 
well known feature of turbulent MHD, which has been observed in the 
solar wind (where ${\bf v} \approx \pm {\bf B}$). It has been linked 
with a principle of minimum energy at constant cross-helicity (see 
chapter 7.3 of \cite{Biskamp93} and references therein). This feature 
is also present in numericals simulation of decaying 2D MHD 
turbulence, where the current and the vorticity are seen to be very 
much equal \cite{Kinney98}. This can therefore be seen as the mere 
outcome of conservation of quadratic integral of motions, and may 
provide an interesting general rule about dynamo saturation in 
systems where these quadratic constraints are dominant.

Using the Gaussian shape for the fluctuations, it is quite easy to derive the mean properties of the fluctuations. To do so, we will make use of the following standard results \cite{Lumley70}:
\EQ
Z = (2 \pi)^3 \sqrt{\det[A]} = (2 \pi)^3 [\beta^2 - \mu_c^2]^{3/2} \; , \qquad \langle x_i x_j \rangle = (A^{-1})_{ij} \; .
\EN 
Then, it is easy to show that part of the energy is going into the fluctuations and that there is equipartition between the fluctuating parts of the magnetic energy and of the kinetic energy:
\EQ
\langle u'^2 \rangle = \langle b'^2 \rangle = \frac{3\beta}{\beta^2 - \mu_c^2} \; .
\EN
One can also calculate the quantity of cross helicity going into the fluctuations:
\EQ
\langle \vec{u'} \cdot \vec{b'} \rangle = - \frac{3\mu_c}{\beta^2 - \mu_c^2} \; .
\EN
One should notice that there is no net magnetic helicity in the fluctuations because of the fact that $A$ is strictly conserved. Then, the fractions of magnetic energy, cross helicity and kinetic energy going into the fluctuations are:
\EQA
\frac{\langle b'^2 \rangle}{\int \bar{B}^2 \ d{\bf x}} &=& \frac{\langle {\bf u' \cdot b'} \rangle}{\int \bar{U} \cdot \bar{B} \ d{\bf x}}  = \frac{3 \beta}{\beta^2-\mu_c^2} \mathcal{M}^{-1} \; , \\ \nonumber
\frac{\langle u'^2 \rangle}{\int \bar{U}^2 \ d{\bf x}} &=& \frac{\beta^2}{\mu_c^2}  \, \frac{3 \beta}{\beta^2-\mu_c^2} \mathcal{M}^{-1} \; ,  
\ENA 
where $\mathcal{M} = \int \bar{B}^2 \ d{\bf x}$ is the magnetic energy
of the coarsed-grained field. The first equation shows that there is
an equal fraction of magnetic energy and cross helicity which goes in
the fluctuations and the positivity of the magnetic energy requires:
$\beta^2 > \mu_c^2$. Using this inequality and the second line, we can
show that the fraction of kinetic {\bf energy} going into the
fluctuations is then bigger than that of the magnetic energy and cross
helicity. This may gives some mathematical ground to the energy
minimization procedure we used in section \ref{EnergyMethod}.

\subsubsection{Rotating case}
\label{Rotating}
The situation is changed when the other constant of motion are taken 
into account. We have:
\EQA
\delta H_{m}' &=& 2\int \frac{b}{r}\ \delta\rho\ d{\bf x} d{\bf u}d{\bf b} \, ,
\\ \nonumber
\delta L' &=& \int u r\ \delta\rho\ d{\bf x} d{\bf u}d{\bf b} \, .
\ENA
On the other hand,
\EQA
\delta L = \int (\delta\bar{A}\ \bar{U}+\bar{A}\ \delta \bar{U})\ 
r^{2}\ d{\bf x} &=& \int (\bar{U}\ {\bf curl}^{-1}\delta \bar{\bf 
B}_{P}+\bar{A}\ \delta \bar{U}) \ r^{2} d{\bf x} \\ \nonumber
&=& \int ({\bf curl}^{-1}(r^{2}\bar{U})\cdot \delta \bar{\bf 
B}_{P}+\bar{A}\ \delta \bar{U}\ r^{2})\ d{\bf x} \\ \nonumber
&=& \int ({\bf curl}^{-1}(r^{2}\bar{U})\cdot {\bf b}_{P}+\bar{A} \ u 
\ r^{2})\ \delta\rho \ d{\bf x} d{\bf u}d{\bf b} \, .
\ENA
Adding Lagrange multipliers $-\mu'_m$, $-\gamma$ and $-\gamma'$ for 
$H_{m}'$, $L$ and $L'$ respectively, we find that the expression 
(\ref{g1}) is multiplied by
\EQ
{\rm exp}\biggl\lbrace -2\mu'_m \frac{b}{r}-\gamma' \ r \ u-\gamma \ 
({\bf curl}^{-1}(r^{2}\bar{U})\cdot {\bf b}_{P}+\bar{A} \ u \ 
r^{2})\biggr\rbrace \, .
\EN
The distribution of fluctuations is then still Gaussian and given by 
(\ref{g2}) but now the mean-field equations are
\begin{eqnarray}
\beta {\bf U}_{P}+\mu_{c}{\bf B}_{P}={\bf 0} \, , \\ \nonumber
\beta {U}+\mu_{c}{B}+\gamma' r+\gamma \bar{A} r^{2}={0} \, , \\ \nonumber
\beta B+2\mu_{m}A+\mu_{c}U+\frac{2\mu'_m}{r}=0 \, , \\ \nonumber
\beta{\bf B}_{P}+2\mu_{m}{\bf A}_{P}+\mu_{c}{\bf U}_{P}+{\bf 
curl}^{-1}\lbrack rC'(rA)\rbrack+\gamma \ {\bf 
curl}^{-1}(r^{2}U)={\bf 0} \, .
\end{eqnarray}
Taking the ${\rm curl}$ of the vectorial relations, we get the system 
(\ref{Equilibr2}).\

\begin{figure}
\begin{center}
\includegraphics[scale=1,clip]{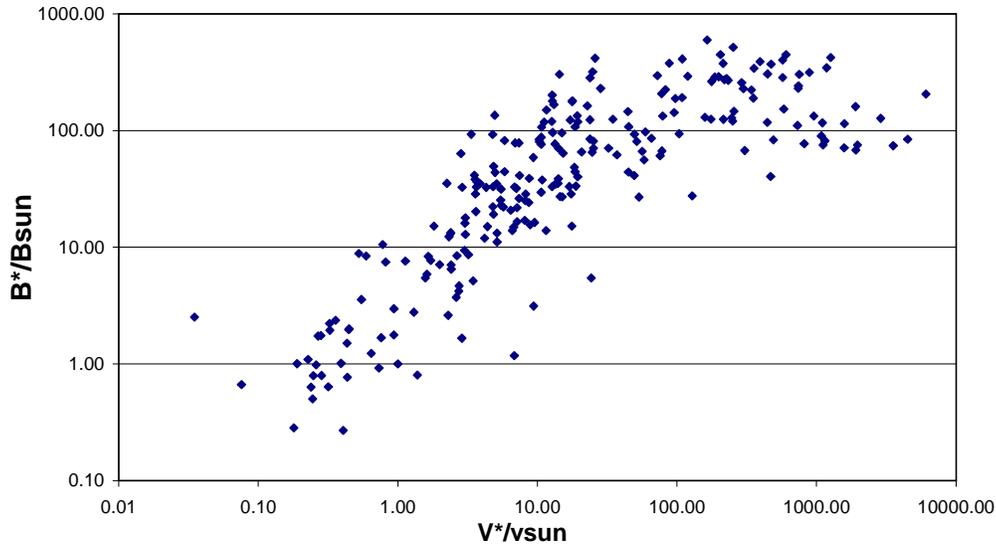}
\caption{\label{Etoiles} Magnetic field of stars (of late-type dwarfs) calculated from their X-ray emission, versus their rotation velocity.}
\end{center}
\end{figure}

Therefore, in the pesent case taking into account additional constant 
of motions, the relation between the velocity and the magnetic field 
is not linear anymore. The linearity is only valid for the poloidal 
component. The toroidal component obeys:
\begin{eqnarray}
\beta (U+ \frac{\gamma'}{\beta} r)= -\mu_{c}B-\gamma A r^{2} \, .
\end{eqnarray}
We can interprete $U+ \gamma' / \beta r$ as the relative velocity
around a solid rotation $\Omega=-{\gamma'/ \beta}$. Indeed, $\gamma'$
is the Lagrange multiplier for the angular momentum constraint. The
non-trivial term responsible for the departure from linearity is
$-\gamma A r^{2}$. Thus, the breaking of the proportionnality between
the velocity and the magnetic field can be attributed to the
conservation of the angular momentum in the Chandrasekhar model. This
is an interesting feature because this conservation rule is likely to
be more relevent in rapidly rotating objects. This may explain the
dynamo saturation in rotating stars, where linearity between magnetic
and velocity field is observed for slowly rotating stars and is broken
for rotator faster than a certain limit  (cf figure \ref{Etoiles}). However, the non-proportionnality between
velocity and magnetic field can also be due to additional conserved
quantities such as those considered by Woltjer \cite{Woltjer59}.

\section{Summary}
\label{Conclusion}

We have developped a statistical theory of axisymetric MHD equations
generalizing the 2D approach by \cite{Jordan97}. We derived the
velocity and magnetic field distribution, and computed the
corresponding equilibrium profiles for the mean flow. Like in the 2D
case, the fluctuations around the mean field are found Gaussian, an
universal feature connected to the conservation of the Casimirs under
the coarse-graining. The equilibrium profiles are characterized by an
alignment of the velocity and magnetic field, which is broken when the
angular momentum conservation is taken into account. The statistical
equilibrium profiles are found to correspond to profiles obtained
under minimization of energy subject to the constraints. Thus, in
the MHD case, in the presence of a coarse-graining (or a small
viscosity), the energy is dissipated while the Helicity, the angular
momentum {\it and} the Casimirs are approximately conserved
(hydromagnetic selective decay). In particular, $\overline{E}= \frac{1}{2}\int \overline{{U}^{2}+{B}^{2}}d{\bf x}\neq \frac{1}{2}\int (\bar{U}^{2}+\bar{B}^{2})d{\bf x}$ because part of energy goes into fine grained fluctuations $E_{fluct}=\overline{E}-E_{m.f.}$. Therefore, the metaequilibrium state
minimizes $E$ at fixed $I$, $H_{m}$, $H_{c}$ and $L$. This can be justified in the ``classical case" (section \ref{Classical}) where we showed that the fraction of kinetic energy going into the fluctuating part of the fields was higher than that of the other quantities, namely the magnetic energy and the cross-helicity. The ``rotating case" (section \ref{Rotating}) requires more algebra and is left for further study.

In contrast, in the 2D hydrodynamical case, the Casimirs are fragile
quantities (because they are expressed as function of the vorticity
which is not an integral quantity as the magnetic potential is) and
thus are altered by the coarse-graining procedure. This is true in
particular for a special class of Casimirs $H=-\int C(\omega)d{\bf
x}$, called $H$-functions, constructed with a convex function $C$ such
that $C''>0$. This leads to two very different behaviors of
hydrodynamical turbulence compared to the hydromagnetic one. First,
the H-functions calculated with the coarse-grained vorticity
$\overline{\omega}$ increase with time while the circulation and
energy are approximately conserved (hydrodynamic selective
decay). Thus, the metaequilibrium state maximizes one of the
$H$-functions at fixed $E$ and $\Gamma$. For example, Chavanis and
Sommeria \cite{jfm1} showed that in the limit of strong mixing (or for
gaussian fluctuations), the quantity to maximize is minus the
enstrophy, giving some mathematical basis to an (inviscid) ``minimum
enstrophy principle". In this context, $\overline{\Gamma}_{2}=\int\overline{\omega^{2}}d{\bf x}\neq \int\bar{\omega}^{2}d{\bf x}$ because part of enstrophy goes into fine-grained fluctuations $\Gamma_{fluct}=\bar{\Gamma}_{2}-\Gamma_{2}^{m.f.}$. However, for more general situations, the
$H$-function that is maximized at metaequilibrium is non-universal and
can take a wide diversity of forms as discussed by Chavanis
\cite{Chavanis03}.  Due to their resemblance with entropy functionals
(they increase with time, one is maximum at metaequilibrium,...), and because they
generally differ from the Boltzmann entropy $S_{B}=-\int \omega \ln
\omega d{\bf x}$, the $H$-functions are sometimes called ``generalized
entropies''
\cite{Chavanis03}.  From the statistical mechanics point of view,
there is an infinite number of constraints (depending on the micro
scale fields) to take into acccount when deriving the Gibbs
state. Consequently, the shape of the fluctuations is not
universal. This is why the $H$-function that is maximized at
metaequilibrium is also non-universal. However, if the distribution of
fluctuations is imposed by some external mechanism (e.g., a
small-scale forcing) as suggested by Ellis et al. \cite{ellis}, the
functional $S[\overline{\omega}]$ is now a well-determined functional
determined by the Gibbs state and the prior vorticity distribution
\cite{ellis,Chavanis03}.

Our computation can provide interesting insight regarding dynamo 
saturation. It is however limited by its neglect of dissipation and 
forcing mechanism. It would therefore be interesting to generalize 
this kind of approach to more realistic systems. In that case, the 
entropy might not be the relevent quantity anymore, but rather the 
turbulent transport, or the entropy production \cite{Dewar03}.

\appendix*
\section{Curl operators}
\label{AnnexeA}
Following Jordan and Turkington, we define
\EQA
{\rm curl}{\bf B}&=&(\nabla\times {\bf B}) \cdot {\bf e}_{\theta}
\\ \nonumber
{\bf Curl}A &=& \nabla\times (A \, {\bf e}_{\theta})
\ENA
for any vector ${\bf B}$ and scalar $A$. It is straightforward to 
show that we have the following relations:
\EQA
{\rm curl}{\bf Curl}(\frac{A}{r}) &=& - r \Delta_{*} A \\
\int A \ {\rm curl}{\bf B} \ d{\bf x} &=& \int {\bf Curl}A \cdot {\bf 
B}\ d{\bf x}
\ENA
Setting $A={\rm Curl}^{-1}{\bf B}'$ and ${\rm curl}{\bf B}=A'$ in the 
last identity, we get
\EQ
\int {\rm Curl}^{-1}{\bf B}' \ A' \ d{\bf x} = \int {\bf B}'\cdot 
{\bf curl}^{-1}A' \ d{\bf x}
\EN

\bibliographystyle{plain}

\end{document}